# An integrated Micromegas UV-photon detector

Joost Melai[a*], Alexey Lyashenko[b], Amos Breskin[b], Harry van der Graaf[c], Jan Timmermans[c], Jan Visschers[c], Cora Salm[a], Jurriaan Schmitz[a]

[a] *University of Twente/Mesa+ Institute for Nanotechnology, Hogekamp 3214, PO box 217, Enschede 7500 AE, The Netherlands*
[b] *Department of Particle Physics, Weizmann Institute of Science, Rehovot, Israel*
[c] *NIKHEF, Amsterdam, the Netherlands*

**Abstract**

Preliminary results of a photon detector combining a Micromegas like multiplier coated with a UV-sensitive CsI photocathode are described. The multiplier is made in a CMOS compatible InGrid technology, which allows to post-process it directly on the surface of an imaging IC. This method is aimed at building light-sensitive imaging detectors where all elements are monolithically integrated. We show that the CsI photocathode deposited in the InGrid mesh does not alter the device performance. Maximum gains of ~6000 were reached in a single-grid element operated in Ar/CH4, with a 2% Ion Back Flow fraction returning to the photocathode.

## 1. Introduction

Gaseous photomultipliers [1] with single-photon sensitivity have been playing a major role in Ring Imaging Cherenkov detectors. Following MWPCs with CsI-coated photocathode planes, more advanced photon imaging detectors were developed, like CsI-coated cascaded-GEMs (Gas Electron Multipliers), cascaded GEM and ion-blocking Microhole & Strip Plates (MHSP) elements [2] and cascaded-THGEMs (Thick GEM) [3]. The new photon detectors with micropattern gaseous multipliers present considerable advantages over MWPCs, mostly in terms of higher gain and reduced secondary effects [1].

In this work we investigated the possibility of integrating CsI-photocathodes with the new InGrid technology as presented in [4]. The concept is somewhat similar to the work presented in [5], combining a Micromegas multiplier with a reflective photocathode.

InGrid is an integrated gaseous detector designed to be post-processed directly onto the surface of any CMOS chip; most interestingly it can be built on imaging chips such as Timepix [6]. The structures consist of a punctured metal foil (similar to a Micromegas structure) supported by insulating pillars that keep the grid at a distance of 50–80 μm from the anode plane (figure 1).

The final goal is to create a single-photon sensitive imaging detector by monolithic integration of a readout chip (e.g. Timepix), the Micromegas-like electron multiplier and a photocathode.

There are several important elements to be considered for the successful design and operation of these detectors.

The efficiency with which photoelectrons are extracted from the surface and with which they are then collected into the holes of the multiplier depends on the gas, on the holes geometry and on the electric field distribution. The gain is naturally important for reaching high single-photon sensitivity; it is often limited by photon- and ion-induced secondary (feedback) effects, as reviewed in [1]. The compatibility of the multiplier with the photocathode material and its deposition process are not to be overlooked.

## 2. Experimental

The detector structures were made in the MESA+ cleanroom facility in the University of Twente, the Netherlands. The devices were then transported to the Weizmann Institute in Israel where the photocathodes were deposited, followed by systematic detector investigations.

*2.1. Detector manufacturing*

The technology used to fabricate these devices is entirely compatible with underlying CMOS devices. Most importantly in this respect is the maximum temperature during processing. To preserve CMOS interconnect integrity the temperature should not exceed 400–450 °C. The temperature needed to fabricate the SU-8 polymer support structure does not exceed 95 °C. The sputtering process needed for metal deposition has also been found to be compatible with the underlying chips [4].

The detectors presented in this work did not contain any CMOS devices. They were built on top of dummy chips containing an unpatterned single anode made out of sputtered Al on silicon. Onto this, a layer of SU-8 is deposited by spincoating. SU-8 is a photo-lithographically patternable polymer. This polymer is imaged to define the locations of pillars using a masked exposure. On top of the exposed and partially cross-linked resist layer the grid metal (1 μm Al) is deposited by sputtering. The grid metal is patterned using conventional photolithography and etching processes. After that the SU-8 layer is developed using acetone and isopropyl alcohol. The non-crosslinked parts of the SU-8 are washed away from underneath the metal grid, leaving the cross-linked pillars to support it. The process sequence and an example of the finished structure are shown in figure 1. A more detailed description of the process can be found in [4].

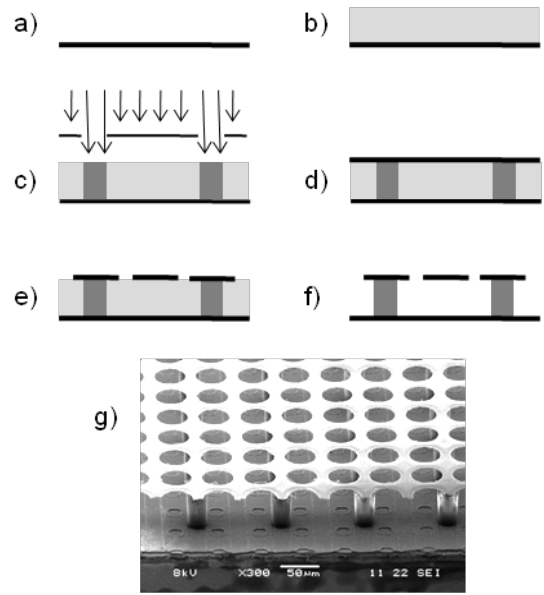

Figure 1: a)–f) Process sequence to create InGrid detectors and g) SEM image of the final structure. The sequence is a) preparation of Al anode, b) coating of SU-8 polymer layer, c) photoimaging of pillar regions using a masked exposure, d) deposition of Al grid material by sputtering, e) patterning of the grid, f) development of the SU-8 layer to create a suspended grid.

The presented study was carried out with small circular InGrid detector prototypes of 20 mm diameter. There was a continuous SU-8 polymer wall around the edge of the detector to prevent discharges. The active inner part of the detector was 18 mm in diameter. The holes in the metal grid had a diameter of 13 μm and a pitch of 20 μm. The grid-to-anode amplification gap was defined by the height of the SU-8 pillars; for present detectors it was kept at 80 μm.

The grid was coated with a 200 nm thick reflective CsI photocathode (PC), by thermal evaporation. For this the devices were placed in a vacuum chamber; the CsI was evaporated from a solid source, kept at a distance of ~40 cm from the InGrid device, to limit the detector's thermal budget. No detrimental effects were observed.

*2.2. Measurement method*

The structures were placed inside stainless steel vacuum vessels, evacuated to $10^{-5}$ mbar prior to gas filling. The chamber was flushed continuously either with Ar/CH$_4$ (95/5) or with Ne/CH$_4$ (at different ratios). The outlet was connected through an oil-bubbler to prevent contamination, keeping the operation pressure just above atmospheric. The UV-photon detector is shown in figure 2. A mesh cathode was placed 5 mm above the grid, biased at the grid's potential. This created a dipole field in the region just above the grid; the electric field being perpendicular to the surface and then bending towards the nearest hole. The mesh cathode was kept at the same potential as the grid to ensure that all field lines extend through the holes into the amplification region below the grid. The field distribution served two purposes; to extract the primary photoelectrons from the photocathode and to guide them towards the holes.

The primary electrons collected into the multiplication region are multiplied by the high field strength below the grid (e.g. 40–60 kV/cm). The resulting charge cloud was collected at the anode and was recorded with external electronics. The layout and operation mechanism of this detector are shown in figure 2.

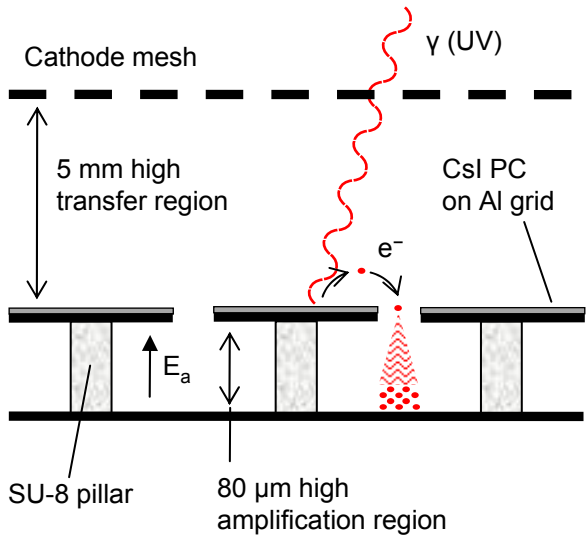

Figure 2: Schematic view and operation principle of the InGrid/CsI photon detector. The region above the grid is used for extraction and transfer of photoelectrons; the part below the grid is used for multiplication of the primary charge carriers.

## 3. Results and discussion

The photocathode was characterized by measuring the primary photocurrent under UV illumination with an Ar(Hg) lamp. In these measurements, the mesh cathode was positively biased and the photocurrent was measured on the grounded grid. The result is shown in figure 3. An electric field strength of ~0.5 kV/cm above the grid was sufficient for reaching a plateau in photoelectron extraction from the photocathode. The monotonic increase in photocurrent is similar to earlier observations with CsI photocathodes. Finite-element simulations [7] indicated that for our device geometry and operation conditions the minimum field strength above the grid, at the surface between the holes in the grid, is at least 2 kV/cm. For very small holes the field in between the holes drops below the 0.5 kV/cm limit and the overall extraction efficiency is lowered. If the holes are too large the efficiency decreases due to loss of effective photocathode surface area.

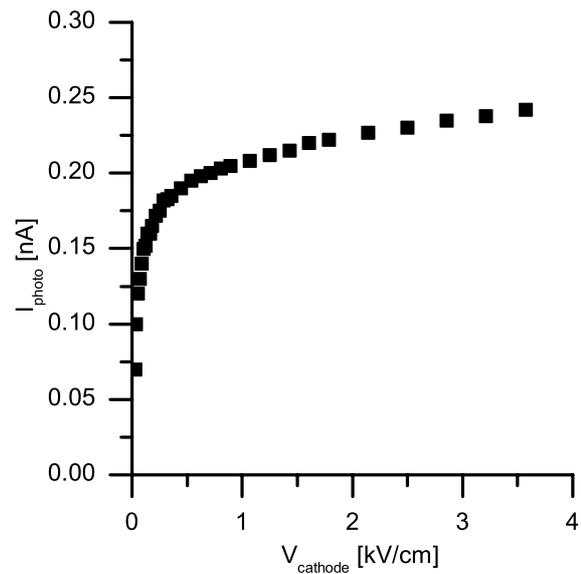

Figure 3: Photocurrent extraction from the grid of the detector shown in figure 2, measured in Ar/CH4 (95/5).

Figure 4 shows gain curves obtained under UV illumination in the detector of figure 2. For these measurements the anode was grounded, the grid and the mesh cathode above were biased at the same negative potential. Gain is defined here as the ratio of the anode current to the plateau current before the onset of multiplication. Gains of the order of 6000 were measured before discharge limit. Best results were obtained in Ar/CH$_4$ (95/5). Note the slight deviation from the typical exponential behavior, observed in Ne and in Ne/CH$_4$ (98/2), due to the onset of secondary effects.

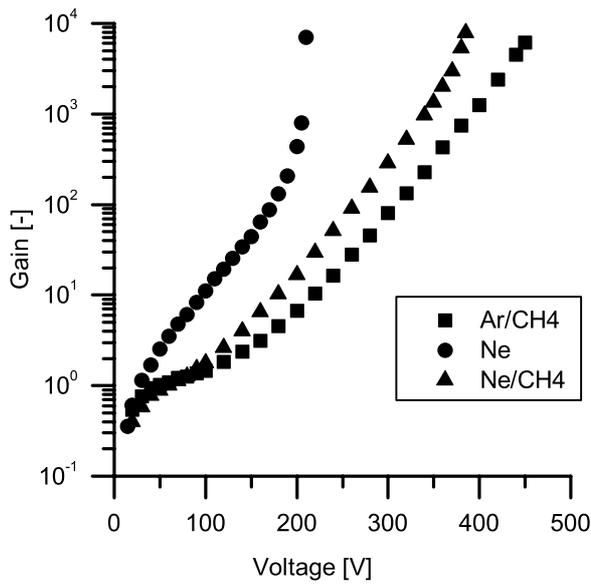

Figure 4: Gain curves measured under UV illumination in different gas mixtures, obtained from current measurements on the grounded anode, the grid and mesh cathode were kept at the same negative HV.

The current level at low bias, where gain is unity, corresponds very well to the saturation level of the photoelectron extraction in figure 3. This suggests that the transfer and collection efficiency of extracted primary electrons into the holes of the grid is very high.

Figure 5 shows a comparison between two gain measurements on the same device. First the uncoated device was characterized placing a thin (10 nm) semitransparent photocathode deposited onto a transparent window, above the grid; it was kept at ground potential, while the grid and the anode were positively biased in such a way to allow for efficient photoelectron extraction (field above the grid > 0.5 kV/cm) and charge multiplication (high field below the grid). In a second step, the semitransparent photocathode was removed and the detector was operated with a reflective photocathode deposited on the grid. The gain behavior of the detector was found to be similar in both cases, as shown in figure 5; this shows that the InGrid detector operation was not affected by CsI photocathode deposition onto the grid surface. At the same time the CsI is compatible with the materials of the InGrid detector.

Current measurements on all three electrodes allowed assessing the fraction of avalanche ions returning upwards to the cathode. This fraction is known as the avalanche induced Ion Backflow Fraction (IBF) [2,8]. It is defined as the fraction of the rise in cathode current (in reference to the cathode current at 0 V amplification bias) compared to the anode current. This measurement was performed using the semi-transparent photocathode. The result is shown in figure 6; IBF reached a level of ~2% at a gain of 4000.

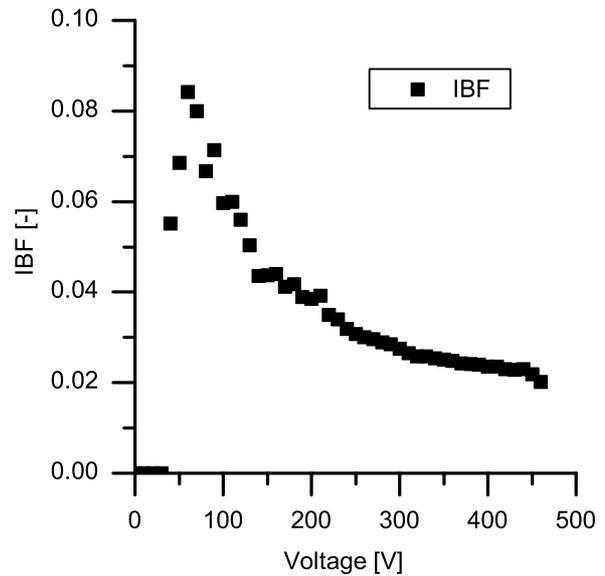

Figure 6: IBF measured in Ar/CH4, calculated from the fractional increase of cathode current with respect to the anode current, in the detector depicted in figure 2.

Better ion-blocking and higher gains are expected with multi-grid structures [9]. These could be useful for possible applications to visible-light imaging [10,11] with bialkali-coated InGrid devices.

## 4. Conclusions

We have presented first results of a UV-sensitive photon detector, based on the Ingrid technology, having a CsI-coated grid photocathode. The measurements indicated good photoelectron extraction from the photocathode and collection into the multiplication gap; ion backflow fraction values of the order of 2% were measured at gains of 4000.

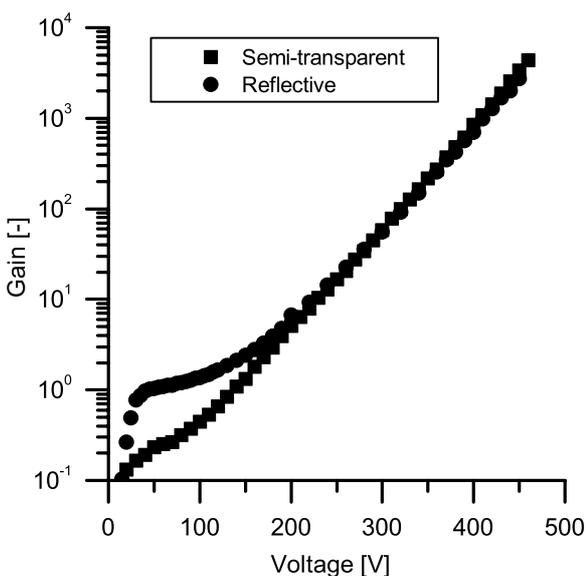

Figure 5: Comparison of gain in Ar/CH4 (95/5) using a device with a reflective photocathode and the same device before coating coupled to a semi-transparent photocathode located above the grid.

These gains were insufficient for single-photoelectron pulse counting. New devices built on Timepix chips are in preparation; with their individual-pixel readout electronics we expect to perform high sensitivity single-photon imaging.

## Acknowledgements

The authors acknowledge Victor Blanco Carballo, Sander Smits, Yevgen Bilevych, Martin Fransen and Moshe Klin for their help during the manufacturing of the test devices and the measurements. This research is funded by Dutch technology foundation STW through project TET-6630. A. Breskin is the W.P. Reuther Professor of Research in The Peaceful Use of Atomic Energy.